\documentclass[amsmath,trackchanges]{aastex702}

\begin{document}

\title{Where Do Population~III Stars Form?
Unveiling Late PopIII-Forming Environments with CampFIRE-6k}

\author[0009-0007-7054-6470]{Elizabeth Bryce}
\affiliation{Department of Physics, Emory University, Atlanta, GA 30322, USA}
\email[show]{eebryce@emory.edu}  

\author[0000-0003-2237-0777]{Alessandra Venditti}
\altaffiliation{Cosmic Frontier Center Prize Fellow}
\email{alessandra.venditti@utexas.edu}
\affiliation{Department of Astronomy, University of Texas at Austin, 2515 Speedway, Stop C1400, Austin, TX 78712, USA}
\affiliation{Cosmic Frontier Center, The University of Texas at Austin, Austin, TX 78712}

\author[0000-0002-8429-4100]{Jenna Samuel}
\affiliation{Department of Astronomy, University of Texas at Austin, 2515 Speedway, Stop C1400, Austin, TX 78712, USA}
\affiliation{Cosmic Frontier Center, The University of Texas at Austin, Austin, TX 78712}
\email{jenna.samuel@austin.utexas.edu}

\author[0000-0003-0212-2979]{Volker Bromm}
\affiliation{Department of Astronomy, University of Texas at Austin, 2515 Speedway, Stop C1400, Austin, TX 78712, USA}
\affiliation{Cosmic Frontier Center, The University of Texas at Austin, Austin, TX 78712}
\affiliation{Weinberg Institute for Theoretical Physics, University of Texas at Austin, Austin, TX 78712, USA}
\email{vbromm@astro.as.utexas.edu}

\author[0000-0002-8984-0465]{Julian B. Muñoz}
\affiliation{Department of Astronomy, University of Texas at Austin, 2515 Speedway, Stop C1400, Austin, TX 78712, USA}
\affiliation{Cosmic Frontier Center, The University of Texas at Austin, Austin, TX 78712}
\affiliation{Weinberg Institute for Theoretical Physics, University of Texas at Austin, Austin, TX 78712, USA}
\email{julianbmunoz@utexas.edu}

\begin{abstract}

Population III (PopIII) stars are the first generation of metal-free stars, thought to form in minihalos ($M_\mathrm{halo} \sim10^5-10^6\,\mathrm{M_{\odot}}$) at Cosmic Dawn ($z\gtrsim15$) before largely giving way to Population II stars formed from the newly enriched medium. Simulations show, however, that PopIII stars may continue forming at low rates through the Epoch of Reionization (EoR, $z \sim 6 - 10$), due to the inhomogeneous enrichment of the Universe. Nonetheless, significant uncertainty remains on the rates and typical environments of late PopIII formation. We use the CampFIRE-6k cosmological zoom-in simulation to examine PopIII star formation history and environments through $z\sim6$. We find that PopIII stars form at a near constant rate ($\sim 10^{-3.5} ~\mathrm{M_{\odot}\,yr^{-1}\,cMpc ^{-3}}$) through $z=6$, including within atomic-cooling halos ($M_\mathrm{halo}\sim 10^7-10^8\,\mathrm{M_\odot}$) at late times.

\end{abstract}




\section{Introduction} 

PopIII stars formed from primordial gas and are expected to be predominantly massive and short-lived, rapidly enriching their surroundings \citep{Bromm2013,Klessen2023}. Direct detection of the earliest PopIII stars is challenging due to their faintness and short-lived nature \citep{Schauer2020}.
However, cosmological simulations predict that chemically pristine gas pockets may survive until $ z \sim 6$, enabling continued PopIII star formation through the EoR \citep{Xu2016, Sarmento2018, Liu2020, Venditti_2023, Sarmento2025}. These late-PopIII sites may be directly observable with the James Webb Space Telescope (JWST) \citep{Venditti2026r}, and candidates have been proposed at redshifts as low as $ z \sim 3$ \citep{Mondal2025, Cai2025}.

The environments hosting late PopIII formation remain nonetheless poorly constrained, providing little to no guidance for observational searches, and leaving gaps in our understanding of early galaxy formation. Late PopIII star formation may occur in underdense regions with a slower global enrichment that preserves the required pristine gas \citep{Trenti2009, Xu2016}. However, theoretical models also support PopIII formation during the EoR near enriched star-forming regions and in atomic-cooling halos (ACHs), substantially more massive than the first minihalos \citep[e.g.][]{Venditti_2023, Storck_2026, Jeong2026}. 
In this work, we examine late PopIII star-forming sites with the CampFIRE-6k simulation.


\section{Methods} \label{sec:style}

The CampFIRE simulations are zoom-ins of the BonFIRE cosmological hydrodynamical simulation \citep{Samuel2026}. CampFIRE-6k has a box size of 5 cMpc and gas mass resolution $m_\text{baryon} \sim 6 \times 10^3 \,\mathrm{M_\odot}$, running to $z 
\sim 6$. The BonFIRE/CampFIRE suite implements state-of-the-art FIRE-3 physics \citep{Hopkins2023}, with physics models and post-processing described in \cite{Samuel2026}, making them ideal laboratories for studying late PopIII star formation with accurate early Universe physics. The simulations additionally include a simple PopIII model with adjusted feedback to account for a top-heavy initial mass function (IMF) below a critical metallicity ($Z_\text{crit} \sim 10^{-5}\, \mathrm{Z_\odot}$).

We analyze halo and star particle data to track star formation histories and environments across CampFIRE-6k, and examine gas density and metallicity in/around PopIII-forming halos at $z \sim 6$.

\section{Results} \label{sec:floats}

\begin{figure*}[ht!]
\includegraphics[width=\linewidth]{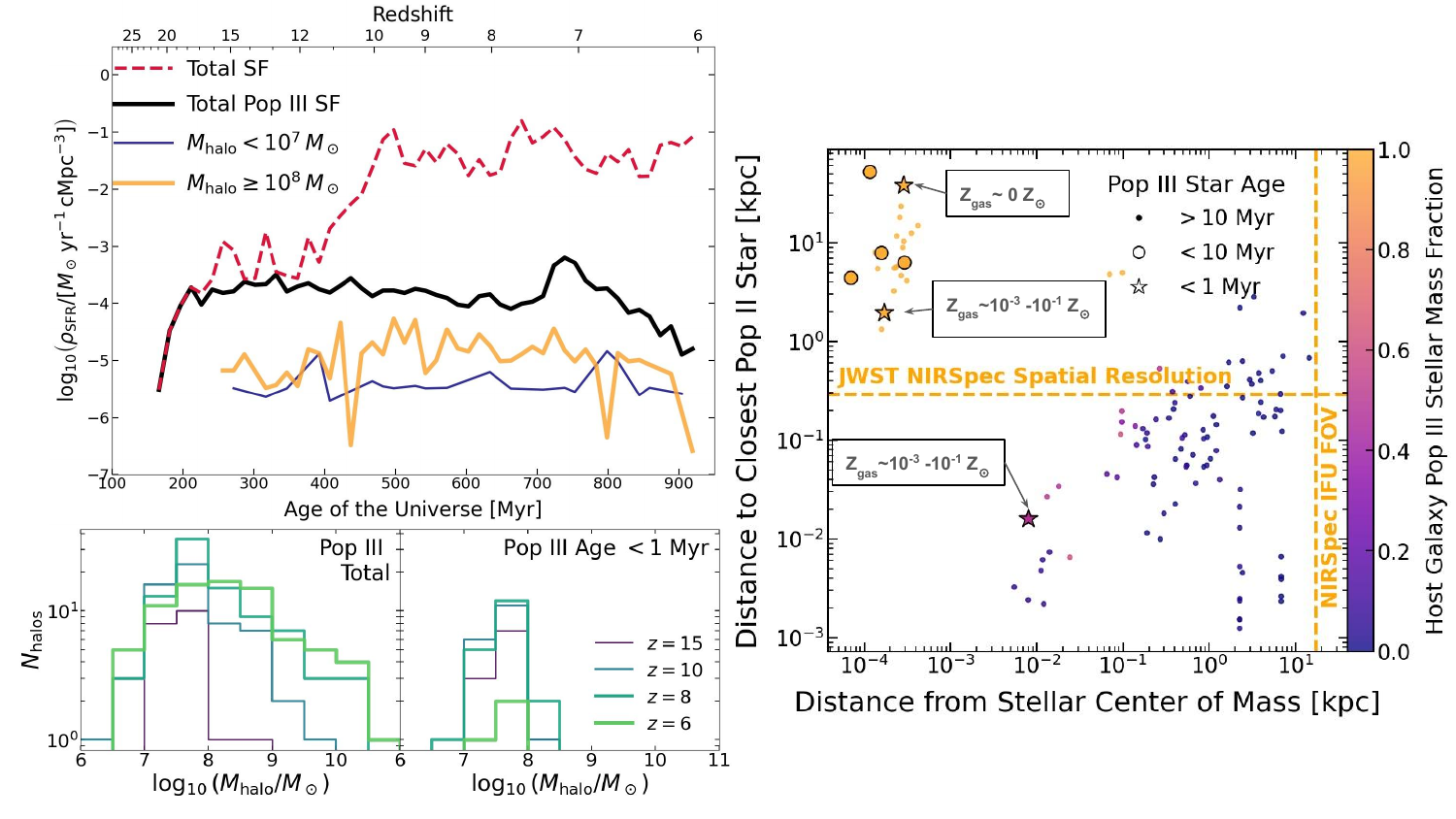}
\caption{\textbf{Top Left}: Total (\textit{thick, dashed}) and PopIII (\textit{thick, solid}) SFRD in CampFIRE-6k, highlighting PopIII contributions from $M_\mathrm{halo} < 10^7 \, \mathrm{M_\odot}$ and $M_\mathrm{halo} > 10^8 \, \mathrm{M_\odot}$ (\textit{thin, solid}), as a function of cosmic time/redshift. \textbf{Bottom Left}: Host halo mass distribution at different redshifts for all PopIII hosts (left panel), and hosts containing PopIII stars younger than 1 Myr (right panel). \textbf{Right}: Distance of PopIII star particles from the nearest PopII star particle vs. from the center of their host halos at $z \sim 6$. PopIII stellar age is represented by \textit{shape}, host galaxy PopIII mass fraction by \textit{color}, and JWST NIRSpec/IFU spatial resolution and field-of-view (FoV) limits are \textit{orange dashed lines}. Metallicity (mass-weighted mean to maximum) within the virial radius is indicated for actively PopIII-forming halos (shown as \textit{stars}). 
\label{fig:general}}
\end{figure*}

While PopII stars begin dominating the total Star Formation Rate Density (SFRD) by $z \sim 12$, PopIII stars maintain a relatively constant rate of $\sim 10^{-4} \,\mathrm{M_{\odot}
\,yr^{-1} \,cMpc ^{-3}}$ from $z \sim 20-7.5$, before a sharp peak up to $~10^{-3} \,\mathrm{M_{\odot}\,yr^{-1} \, cMpc ^{-3}}$, and finally a decrease at $z \lesssim 7$ (Figure~\ref{fig:general}, top left). Nonetheless, at $z \sim 6$ we find seven halos with PopIII stars younger than 10 Myr, three of which are still actively forming PopIII stars (Figure~\ref{fig:general}, right).

PopIII stars in CampFIRE-6k form predominantly in ACHs of $M_\mathrm{halo}\sim 10^7 - 10^8 \, \mathrm{M_{\odot}}$, while older populations also inhabit merged systems with halo masses up to $\sim10^{11} \, \mathrm{M_\odot}$ (Figure~\ref{fig:general}, bottom left).  For example, one PopIII-forming halo at $z \sim 6$ is isolated from enriching sources, with no stars within $\sim 15$ kpc of its young PopIII population, and uniformly pristine composition within the halo's virial radius (Figure~\ref{fig:general}, top right). Conversely, the other two young PopIII stars form in much denser, enriched environments, within  $\sim 5-10$ kpc of other stars and with metallicities up to $0.1 \, \mathrm{Z_\odot}$ within the host virial radii. These PopIII star particles form in pristine clumps of gas, but in close proximity to enriched gas. Thus, late PopIII star formation can occur across substantially different local and large-scale environments.
\section{Discussion and Conclusions} \label{sec:highlight}

Our results are affected by resolution and box size. CampFIRE-6k can accurately resolve halos with masses $\geq 10^{6.5} \, \mathrm{M_{\odot}}$, but cannot capture the minihalo population in detail. Likewise, the 5 cMpc box size limits our ability to capture statistics at the rare/high-mass end. Extending the analysis to the high-resolution CampFIRE-800 ($m_\mathrm{baryon} \sim 800\, \mathrm{M_\odot}$) and the much larger BonFIRE (41.2 cMpc box size) simulations will enable us to robustly assess these biases and connect scales.

Despite these limitations, our analysis indicates that PopIII star formation persists to $z\sim6$ in ACHs at near constant rates, and can even occur near coeval PopII formation. Massive PopIII hosts are especially promising because they could support stronger PopIII starbursts \citep{Jeong2026}, which are more likely to be directly detectable. On the other hand, surviving low-mass PopIII stars with weaker hard-spectral signatures may still leave a distinct chemical fingerprint in higher-mass merged halo systems \citep{Bonifacio2003}.

Further exploration of these unexpectedly diverse halo environments will be integral for our understanding of the drivers of late PopIII star formation, and the identification of promising targets for observational searches.

\begin{acknowledgements}
EB acknowledges support from NSF REU grant
AST-2244278 (PI:Jogee). AV acknowledges funding from the Cosmic Frontier Center and the University of Texas at Austin’s College of Natural Sciences. JS acknowledges support from program JWST-AR-06278. JBM was supported by NSF Grants AST-2307354 and AST-2408637, and by the NSF-Simons AI Institute for Cosmic Origins.
\end{acknowledgements}

%

\bibliography{main}{}
\bibliographystyle{aasjournalv7}



\end{document}